\documentclass[12pt]{article}
\usepackage{epsfig}
\def\be{\begin{equation}}
\def\ee{\end{equation}}

\def\lsim{\raise0.3ex\hbox{$<$\kern-0.75em\raise-1.1ex\hbox{$\sim$}}}
\def\gsim{\raise0.3ex\hbox{$>$\kern-0.75em\raise-1.1ex\hbox{$\sim$}}}

\def\NP{{ Nucl.\ Phys.\ }}

\begin{document}
\thispagestyle{empty}

\vskip 1.5 cm

\centerline{\large{\bf Percolation and Magnetization for}}

\medskip

\centerline{\large{\bf Generalized Continuous Spin Models}}

\vskip 1.0cm

\centerline{\bf Santo Fortunato and Helmut Satz}

\bigskip

\centerline{Fakult\"at f\"ur Physik, Universit\"at Bielefeld}
\par
\centerline{D-33501 Bielefeld, Germany}

\vskip 1.0cm

\noindent

\centerline{\bf Abstract:}

\medskip

For the Ising model, the spin magnetization transition is
equivalent to the percolation transition of Fortuin-Kasteleyn clusters;
this result remains valid also for the 
conventional continuous spin Ising model.
The investigation of more general continuous spin models may help
to obtain a percolation formulation for 
the critical behaviour in $SU(2)$ gauge theory.
We therefore study a broad class of theories, introducing spin
distribution functions, longer range interactions and self-interaction
terms. The thermal behaviour of each model
turns out to be in the Ising universality
class. The corresponding percolation formulations 
are then obtained by extending
the Fortuin-Kasteleyn cluster definition; in several cases
they illustrate recent rigorous results.

\vskip 1cm

\noindent{\bf \large Introduction}

\bigskip

Percolation theory \cite{Stauff,Grimm} has been successfully applied
to describe continuous thermal phase transitions as geometrical
transitions: at the critical threshold, suitably defined clusters
become infinite (see \cite{fisher} - \cite{Cha2}). It is moreover
possible to establish a correspondence between thermal and cluster
variables, such that the critical exponents of corresponding variables
coincide \cite{CK}. This provides an intuitive geometric description of
the mechanism of the thermal transition; e.g., in the case of the Ising
model, we can identify clusters as magnetic domains, so that
magnetization sets
in when they span the lattice. In a recent study \cite{SatzFort}, we have
tried to extend this picture to the confinement-deconfinement transition
of $SU(2)$ gauge theory. This theory leads to a continuous phase transition
which belongs to the universality class of the Ising model, due to the common
$Z(2)$ global symmetry of the corresponding Hamiltonians \cite{Svet}.
However, our result is valid only in the strong coupling limit of $SU(2)$, 
where the partition function can be well approximated \cite{KarGre} by
that of a continuous spin Ising model. In this model, the correspondence
between percolation and thermal variables can be rigorously established
\cite{SanGand,BCG}. For a general study of $SU(2)$ gauge theory, the weak
coupling limit must be taken into account, and there the approximations
used in \cite{KarGre} are no longer valid.

\medskip

To address the deconfinement problem more generally, it may be helpful
to look for an effective theory for $SU(2)$. An interesting attempt in
this direction was proposed some time ago in the search for the fixed
point of the theory by means of block-spin transformations
\cite{okawa}. There the original Polyakov loop configurations were
projected onto configurations of Ising-like spins according to the sign
of the Polyakov loop at each lattice site. This assumes that for the
criticality only the global $Z(2)$ symmetry of the theory is important.
As a result of this projection one obtains an effective Hamiltonian
with only short-range couplings, although some of the resulting
couplings correspond to interactions beyond the fundamental
nearest-neighbour form. Moreover, the assumption that only the $Z(2)$
symmetry of the theory is relevant must be tested.

In this work we investigate several spin models in order to 
get a better understanding of the behaviour of 
the $SU(2)$ effective theory. Our investigations are based
on lattice Monte Carlo simulations of the various models.

We focus on continuous spin Ising models, in which the
individual spins $s_i$ at each lattice site can take on all values in the
finite range $[-1,1]$, with the distribution of the spins governed by
a spin distribution function $f(s_i)$. For $f(s_i)=\delta(|s_i|-1)~\forall
~i$ and only nearest-neighbour (NN) interactions, we recover the
conventional Ising model. For $f(s_i)=1~\forall~i$ and only NN
interactions, we obtain the continuous spin model studied in 
\cite{SanGand}. Here we will consider four more general models of 
this type ($d$ denotes the space dimension):

\medskip

\noindent
A) $d=2$, only NN interactions, but a non-uniform spin distribution
   $f(s_i)$;

\medskip

\noindent
B) $d=2$, $f(s_i)=1~\forall~i$, NN and diagonal
   next-nearest-neighbour (NTN) interaction (Fig.\ 1);

\medskip

\noindent
C) $d=3$, $f(s_i)=1~ \forall~i$, NN and two types of NTN interactions (see
   Fig.\ 2);

\medskip

\noindent
D) as case C), but including an additional self-interaction term
   proportional to $s_i^2 ~ \forall~i$.

\vskip1cm

  \begin{picture}(135,150)
    \put(25,118){\psfig{file=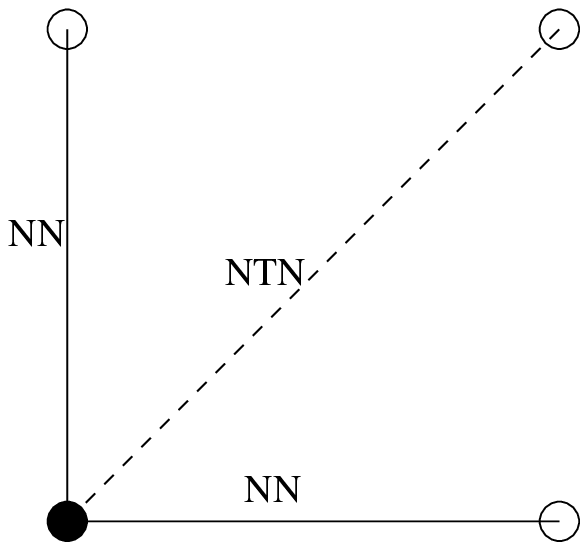,angle=270,width=4cm}}
    \put(20,-10){\begin{minipage}[t]{5cm}{{\footnotesize
          Figure 1. Spin interactions in model B. }}
    \end{minipage}}
    \put(255,14){\epsfig{file=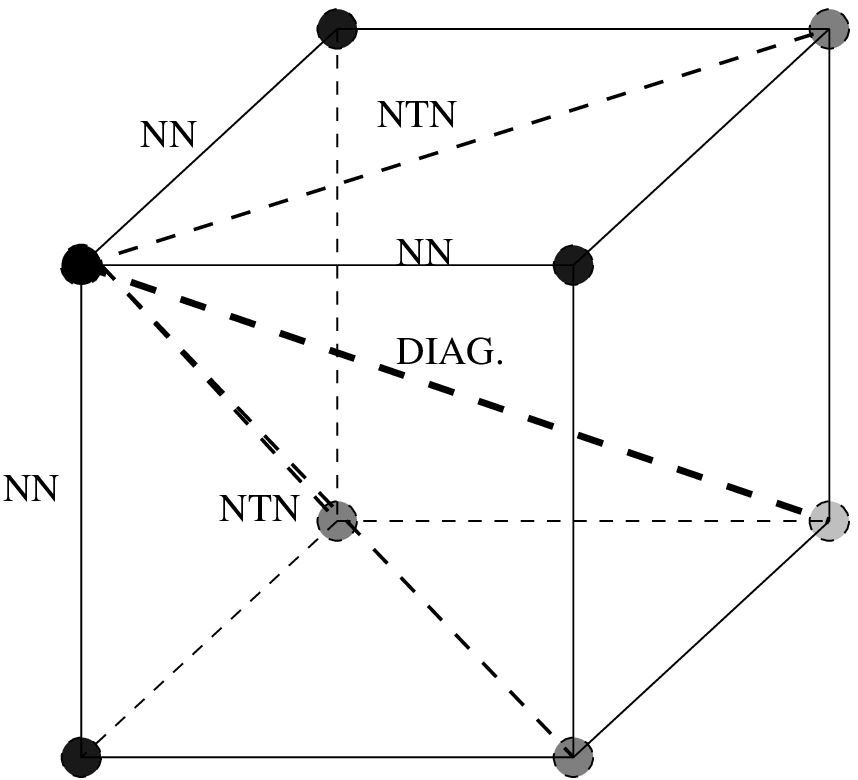,width=4.5cm}}
    \put(255,-10){\begin{minipage}[t]{5cm}{{\footnotesize
          Figure 2. Spin interactions in model C. }}
    \end{minipage}}
  \end{picture}

\vskip2cm

\noindent

We will first present a detailed numerical investigation of 
the thermal critical behaviour and show that each model 
belongs to the same universality class as the Ising model.
Subsequently, we provide a suitable cluster definition
in order to interpret the thermal phase transition 
of the models as a geometrical percolation transition.
Concerning the second issue, it was 
recently proven rigorously  
that the result of \cite{SanGand} can be extended to the models
A, B and C \cite{BCG}. 
We believe, however,
that it is useful to illustrate it by exploiting 
the power of the finite-size scaling analysis.
On the other hand, model D has not been 
investigated so far; as self-interactions play an important
role in gauge theories, it is interesting to examine what happens in this 
case.

\bigskip

\noindent{\bf Model A: Spin Distribution Function}

\bigskip

The spin variable $s$ can take all values in $[-1,1]$. It is convenient
to rewrite it in the form
\be
s=sign(s)\sigma,
\label{1}
\ee
separating the sign from the amplitude of the spin. Since we want to
retain the $Z(2)$ symmetry of the Hamiltonian, the signs of the spins are
equidistributed. In contrast, the amplitudes can in general be weighted
in different ways by choosing a distribution function $f(\sigma)$.
The partition function of such a model then has the form
\be
Z(T) =  \prod_i \int_{0}^{1}
 d\sigma_i f(\sigma_i)~ \exp\{ \kappa
\sum_{\langle i,j \rangle} s_is_j\},
\label{2}
\ee
where $\kappa\equiv J/T$, with $J$ denoting the spin coupling energy and
$T$ the temperature. Since we want to study models with continuous
transitions, the choice
of the function $f(\sigma)$ is not arbitrary. It can be proved that it
must obey certain regularity conditions, which are not very restrictive,
however \cite{ellis}. In \cite{SanGand}, the amplitudes were
equidistributed, with $f(\sigma)=1$. Here we consider
the following form for $f(\sigma)$:
\be
f(\sigma)=\sqrt{1-{\sigma}^2}
\label{3}
\ee
which is the Haar measure of the SU(2) group; it appears in the strong
coupling approximation of SU(2) \cite{KarGre}. To define a cluster, we
use exactly the same procedure as in \cite{SanGand}, i.e., we join
nearest-neighbouring spins of the same sign with the bond probability
$p=1-exp(-2\kappa{\sigma_i}{\sigma_j})$, where $\sigma_i$ and $\sigma_j$
are the amplitudes of the two spins. We use
free boundary conditions
and define percolation when a cluster spans the lattice in each of the
two directions. This is done to exclude the possibility that, due to
the finite size of our lattices, one could find more than one
percolation cluster, which would make the definition of our variables
ambiguous. We shall use this definition for all the models
to be considered here\footnote{In three dimensions even this definition
of a spanning cluster does not exclude the possibility of having more
than one such cluster in the same configuration. Nevertheless the
occurrence of these cases is so small that we can safely ignore them.}.

\medskip

The percolation variables to be studied are: the {\it percolation
strength } $P$,
\be
P={{\it size~of~the~percolating~cluster} \over
            {\it volume~of~the~lattice}}~.  \label{4}
\ee
and the {\it average cluster size } $S$,
\be
S={{\sum_{s}{n_{s}s^2}}\over{\sum_{s}{n_{s}s}}}~.
\label{5}
\ee
Here $n_{s}$ is the number of clusters of size $s$ and the sums exclude
the percolating cluster. Besides these two, we evaluate another very
useful variable. For a given lattice size and a value of $\kappa$,
we determine the relative fraction of configurations
containing a percolating cluster. This variable, that we 
shall call {\it percolation cumulant} and indicate as
$\gamma_r(\kappa)$, is a scaling function, analogous to the Binder
cumulant $g_r(\kappa)$ \cite{Binder}.

\medskip

In addition to these percolation variables, we also measure the lattice
averages of energy and magnetization, in order to study the thermal
transition as well. From the magnetization $M$ we calculate in
particular the physical susceptibility $\chi$,
\be
\chi(\kappa) = V~(\langle{M^2}\rangle-{\langle M \rangle}^2)
\label{6}
\ee
and the Binder cumulant $g_r$,
\be
  g_r(\kappa)=\frac{\langle{M^4}\rangle}{{\langle{M^2}\rangle}^2}~.
\label{7}
\ee
We have performed simulations on four lattice sizes, from $64^2$ to
$240^2$. Our algorithm consists in heatbath steps for the update of the
spin amplitudes followed by Wolff-like cluster updates for the flipping
of the signs. That is basically the same method as used in
\cite{SanGand}, although the heatbath procedure is slightly modified to
take into account the presence of the distribution function
$f(\sigma)$. The proof that the detailed balance condition is satisfied
for such a cluster update is obtained following the derivations
in \cite{Wo2} - \cite{Cha4}. Flipping the spins reduces drastically
the autocorrelation time, which allowed us to collect extensive data and
thus reduce the errors.

\medskip

Fig.\ 3 shows a comparison between the thermal variable
$g_r(\kappa)$ and the percolation cumulant $\gamma_r(\kappa)$, both
as functions of the temperature variable $\kappa$, for different lattice
sizes. The points where the lines cross are the infinite volume
thresholds for the thermal and the geometrical transition,
respectively. The agreement between the two points is remarkable.

\medskip

Since the percolation cumulant is a scaling function, we can
already get an indication about the critical exponents of the
percolation transition. Given the critical point and the exponent
$\nu$, a rescaling of the percolation cumulant as a function
of $(\kappa-\kappa_{cr})L^{1/\nu}$ should give us the same function
for each lattice size ($L$ is the linear
dimension of the lattice). Figs.\ 4 and 5 show the rescaled percolation
cumulant, using $\kappa_{cr}=1.3888$ with the random percolation
exponent $\nu=4/3$ and Ising exponent $\nu=1$, respectively. The
figures clearly show scaling for the Ising exponent and no scaling
for the random percolation exponent.

\medskip

The exponent $\beta$ governs percolation strength and magnetization,
while $\gamma$ governs cluster size and susceptibility. To determine
them, we have performed high-statistics simulations around the critical
point, with the number of measurements for each value of the coupling
varying from 50000 to 100000. We have then used the $\chi^2$ method
\cite{EMSZ} to determine the values of the exponents. The results
are shown in Table I, where the first row shows the calculated
percolation results, the second the corresponding magnetization calculations,
and the third the analytically known Ising model values.
It is evident that the percolation behaviour
coincides fully with the thermal critical behaviour. This conclusion 
holds in general for all admissable distribution functions, as it was shown in
\cite{BCG}.

\bigskip

\noindent{\bf Model B: Next-to-Nearest Neighbour Interactions}

\bigskip

We now want to study how to correctly define clusters in the presence
of more than the standard nearest-neighbour interactions.
We have now two terms, with a Hamiltonian of the form
\be
{\cal H} = -J_{NN} \sum_{\langle i,j \rangle}^{NN} s_is_j -
J_{NTN} \sum_{\langle i,j \rangle}^{NTN} s_is_j
\label{8}
\ee
where the first sum describes nearest-neighbour and the second
diagonal next-to-nearest neighbour interactions (Fig.\ 1). Since longer
range interactions are generally weaker, we fix the ratio
between the two couplings at $J_{NN}/J_{NTN}=10$; however, we do not
believe that our results depend on the choice of the couplings, as long
as both are ferromagnetic. To define clusters, we now extend the
Coniglio-Klein method and define for each two spins $i,j$ of the same
sign, for NN as well as NTN, a bond probability
\be
p_B^x = 1 - \exp \{ - 2 \kappa_x \sigma_i \sigma_j\},
\label{9}
\ee
where $x$ specifies $\kappa_{NN}=J_{NN}/T$ and $\kappa_{NTN}=J_{NTN}/T$,
respectively. 
We recall that in
model B, as well as in C and D, we assume a spin equidistribution,
i.e., $f(\sigma)=1$.

\medskip

We have studied model B using two different Monte Carlo algorithms, in
order to test if a Wolff-type algorithm can also be applied in the 
presence of NTN interactions. The first is the standard Metropolis 
update, while the second alternates
heat bath steps and a generalized Wolff flipping, for which the
clusters are formed taking into account both interactions. The
generalization of the cluster update is trivial. After several runs,
some with high statistics, we found excellent agreement with the
Metropolis results in all cases. So, the mixed algorithm 
with heat bath and Wolff flippings appears to remain viable also
in the presence of more than the standard NN interaction. Subsequently
we have therefore used this mixed algorithm. The lattice sizes 
for model B ranged from $100^2$ to $400^2$.

\medskip

As for model A, we now show for model B the comparison between
percolation cumulant and Binder cumulant, in order to test that the
critical points coincide (Fig.\ 6). We then again rescale the
percolation cumulant, using the critical $\kappa$ determined in Fig.\
6 together with the exponent $\nu$ from the $d=2$ random percolation
transition ($\nu=4/3$) and the $d=2$ Ising model ($\nu=1$).
Figs.\ 7 and 8 indicate that again the correct exponent is that of the
Ising model.

\medskip

The determination of the other two exponents $\beta$ and $\gamma$
by means of the $\chi^2$ method confirms that indeed both the thermal
and the geometrical exponents
belong to the Ising universality class
(Table II).

\bigskip

\noindent{\bf Model C: Extension to Three Dimensions}

\bigskip

We have here
repeated the study for a $d=3$ model with three different interactions
(Fig.\ 2). To fix the model, we have to specify the ratios of the
nearest-neighbour coupling $\kappa_{NN}$ to $\kappa_{NTN}$ and
$\kappa_{diag}$. We chose them to be 10:2 and 10:1, respectively.
Our $d=3$ calculations are performed on lattices ranging from 
$12^3$ to $40^3$.

\medskip

Also here we have first compared the results from a mixed algorithm
of the same kind as for the previous case to those from a standard
Metropolis algorithm; again, the agreement is very good.
Figs.\ 9, 10 and 11 then show the comparison of the thresholds
and the scaling of the percolation cumulant. As before, the
correspondence between percolation and thermal variables is evident
(Table III).

\bigskip

\noindent{\bf Model D: Adding Self-Interaction}

\bigskip

From what we have seen up to now, it seems to be clear that the correct
cluster definition can readily be extended to models including several
(ferromagnetic) spin-spin interactions. However, such terms are not the
only possible interactions in a model with Z(2) symmetry and a
continuous transition. There could be anti-ferromagnetic spin-spin
couplings as well as multispin terms, coupling an even number of
spins greater than two (four, six, etc.). Moreover, since the spins
are continuous, the presence of self-interaction terms is possible,
determined by $s^2$, $s^4$, etc. The treatment for antiferromagnetic
and multispin couplings so far remains an open question. In contrast,
self-interactions are not expected to play a role in the cluster
building, since such terms do not relate different spins. Therefore, we
test a cluster definition ignoring any self-interaction term.

\medskip

We thus consider in Model D the same interactions as in Model C, plus a
term proportional to $J_0 \sum_i s_i^2$. We chose a negative value for
the self-interaction coupling $J_0$; this is the more interesting
case since the corresponding interaction tries to resist the approach
of the system to the ground state at low temperatures ($\sigma=1$
everywhere). The ratios of the NN coupling to the others were
chosen as $J_{NN}:J_{NTN}:J_{diag}:|J_0|= 6:2:1:2$.

\medskip

Again we first verify the viability of the mixed algorithm and then
determine the rescaled percolation cumulant (Fig.\ 12) and the
critical exponents (Table IV). It is evident that the percolation and the
thermal transition again fall into the Ising universality class.

\bigskip

\noindent{\bf Conclusions}

\bigskip

We have performed a complete investigation of several 
continuous spin models, obtained
from the simple model studied in \cite{SanGand} by introducing a distribution function
for the spin amplitudes, longer-ranged interactions and self-interactions.
The thermal behaviour leads in all cases to the Ising universality
class. All systems admit equivalent percolation formulations; 
for models A, B and C, we illustrate numerically the analytical 
results found in \cite{BCG}.
All the features of the models we studied are expected to enter in the 
further extension to SU(2) gauge theories, and the fact that they
do not interfere with a percolation formulation of critical behaviour
provides support for such efforts. 

\bigskip

\noindent{\bf Acknowledgements}

\bigskip

It is a pleasure to thank D.\ Gandolfo for helpful discussions.
We would also like to thank the TMR network ERBFMRX-CT-970122 and 
the DFG Forschergruppe Ka 1198/4-1 for financial support.

\newpage

~~\vskip 2cm

\begin{center}
\begin{tabular}{|c||c|c|c|c|}
\hline
& & & & \\
& $\kappa_{cr}$ & $\beta/\nu$ & $\gamma/\nu$ & $\nu$\\
& & & & \\
\hline
\hline
& & & & \\
Percolation & 1.3887$^{+0.0002}_{-0.0001}$ & 0.128$^{+0.007}_{-0.010}$
& 1.754$^{+0.007}_{-0.008}$ & 0.99$^{+0.03}_{-0.02}$ \\
& & & & \\
\hline
& & & & \\
Magnetization & 1.3888$^{+0.0002}_{-0.0003}$ & 0.121$^{+0.008}_{-0.006}$
& 1.745$^{+0.011}_{-0.007}$
 & 1.01$^{+0.02}_{-0.03}$ \\
& & & & \\
\hline
& & & & \\
Ising Model &                   & 1/8
& 7/4
 & 1\\
& & & & \\
\hline
\end{tabular}\end{center}

\centerline{Table I. Thermal and percolation thresholds and exponents of Model A.}

\vskip 4cm

\begin{center}
\begin{tabular}{|c||c|c|c|c|}
\hline
& & & & \\
& $\kappa_{cr}$ & $\beta/\nu$ & $\gamma/\nu$ & $\nu$\\
& & & & \\
\hline
\hline
& & & & \\
Percolation & 0.9708$^{+0.0002}_{-0.0002}$ & 0.129$^{+0.008}_{-0.009}$
& 1.752$^{+0.009}_{-0.011}$ & 1.005$^{+0.012}_{-0.020}$ \\
& & & & \\
\hline
& & & & \\
Magnetization & 0.9707$^{+0.0003}_{-0.0002}$ & 0.124$^{+0.007}_{-0.005}$
& 1.747$^{+0.009}_{-0.007}$
 & 0.993$^{+0.014}_{-0.010}$ \\
& & & & \\
\hline
& & & & \\
Ising Model &                   & 1/8
& 7/4
 & 1 \\
& & & & \\
\hline
\end{tabular}\end{center}

\centerline{Table II. Thermal and percolation thresholds and exponents of Model B.}

\newpage

~~~\vskip 2cm

\begin{center}
\begin{tabular}{|c||c|c|c|c|}
\hline
& & & & \\
& $\kappa_{cr}$ & $\beta/\nu$ & $\gamma/\nu$ & $\nu$\\
& & & & \\
\hline
\hline
& & & & \\
Percolation & 0.36673$^{+0.00012}_{-0.00010}$ & 0.528$^{+0.012}_{-0.015}$
& 1.9850$^{+0.010}_{-0.015}$ & 0.632$^{+0.01}_{-0.015}$ \\
& & & & \\
\hline
& & & & \\
Magnetization & 0.36677$^{+0.00010}_{-0.00008}$ & 0.530$^{+0.012}_{-0.018}$
& 1.943$^{+0.019}_{-0.008}$
 & 0.640$^{+0.012}_{-0.018}$ \\
& & & & \\
\hline
& & & & \\
Ising Model \cite{ferr}&                   & 0.518$^{+0.007}_{-0.007}$
& 1.970$^{+0.011}_{-0.011}$
 & 0.6289$^{+0.0008}_{-0.0008}$ \\
& & & & \\
\hline
\end{tabular}\end{center}

\centerline{Table III. Thermal and percolation thresholds and exponents of Model C.}

\vskip 4cm

\begin{center}
\begin{tabular}{|c||c|c|c|c|}
\hline
& & & & \\
& $\kappa_{cr}$ & $\beta/\nu$ & $\gamma/\nu$ & $\nu$\\
& & & & \\
\hline
\hline
& & & & \\
Percolation & 0.3005$^{+0.0001}_{-0.0001}$ & 0.524$^{+0.010}_{-0.011}$
& 1.9750$^{+0.008}_{-0.009}$ & 0.636$^{+0.011}_{-0.017}$ \\
& & & & \\
\hline
& & & & \\
Magnetization & 0.3004$^{+0.0002}_{-0.0001}$ & 0.513$^{+0.012}_{-0.010}$
& 1.963$^{+0.014}_{-0.009}$
 & 0.626$^{+0.011}_{-0.010}$ \\
& & & & \\
\hline
& & & & \\
Ising Model \cite{ferr} &                   & 0.518$^{+0.007}_{-0.007}$
& 1.970$^{+0.011}_{-0.011}$
 & 0.6289$^{+0.0008}_{-0.0008}$ \\
& & & & \\
\hline
\end{tabular}\end{center}

\centerline{Table IV. Thermal and percolation thresholds and exponents of Model D.}

\newpage
\begin{picture}(135,360)
\put(80,140){\epsfig{file=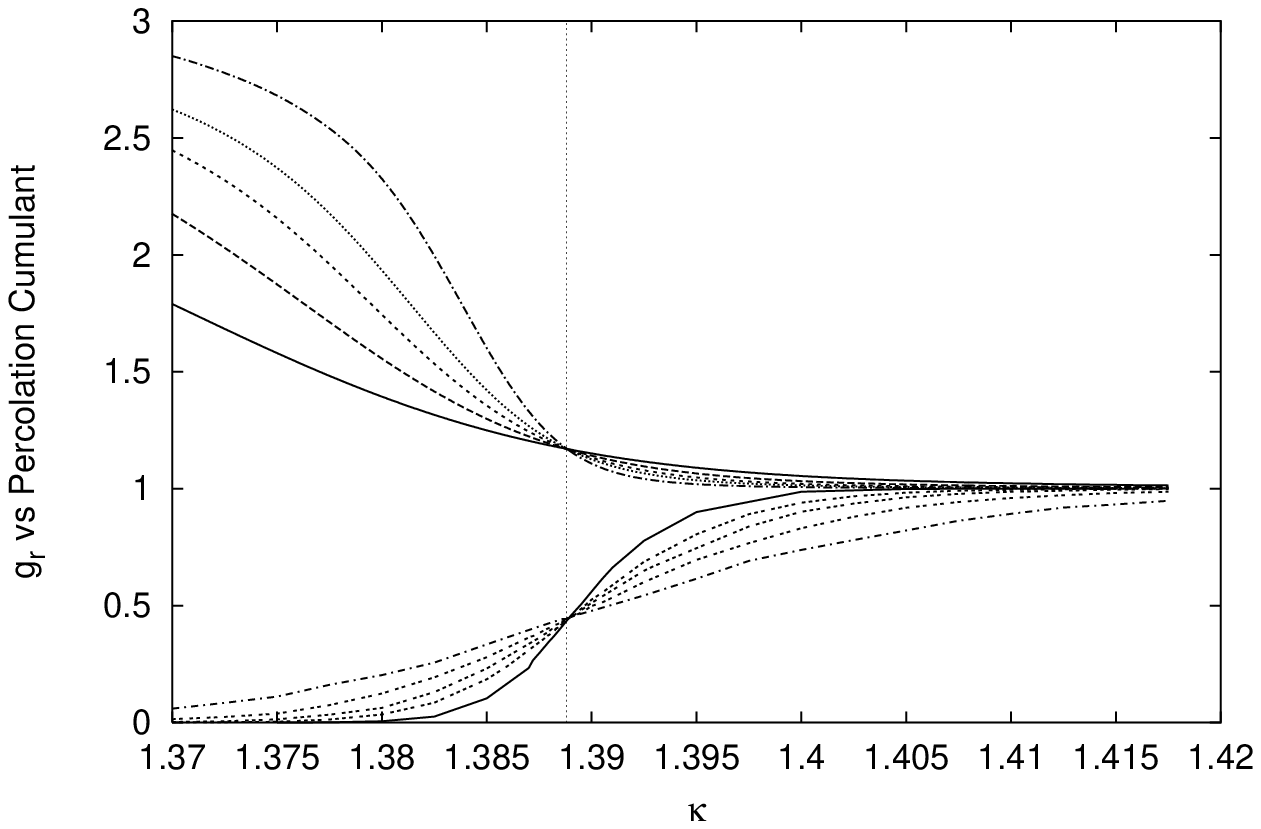,width=9cm}}
    \put(30,130){\begin{minipage}[t]{12cm}{{\footnotesize
          Figure 3. Comparison of the thermal and the geometrical
critical point for Model A obtained respectively from the Binder cumulant
$g_r$ and the percolation cumulant.}}
\end{minipage}}
\put(80,-70){\epsfig{file=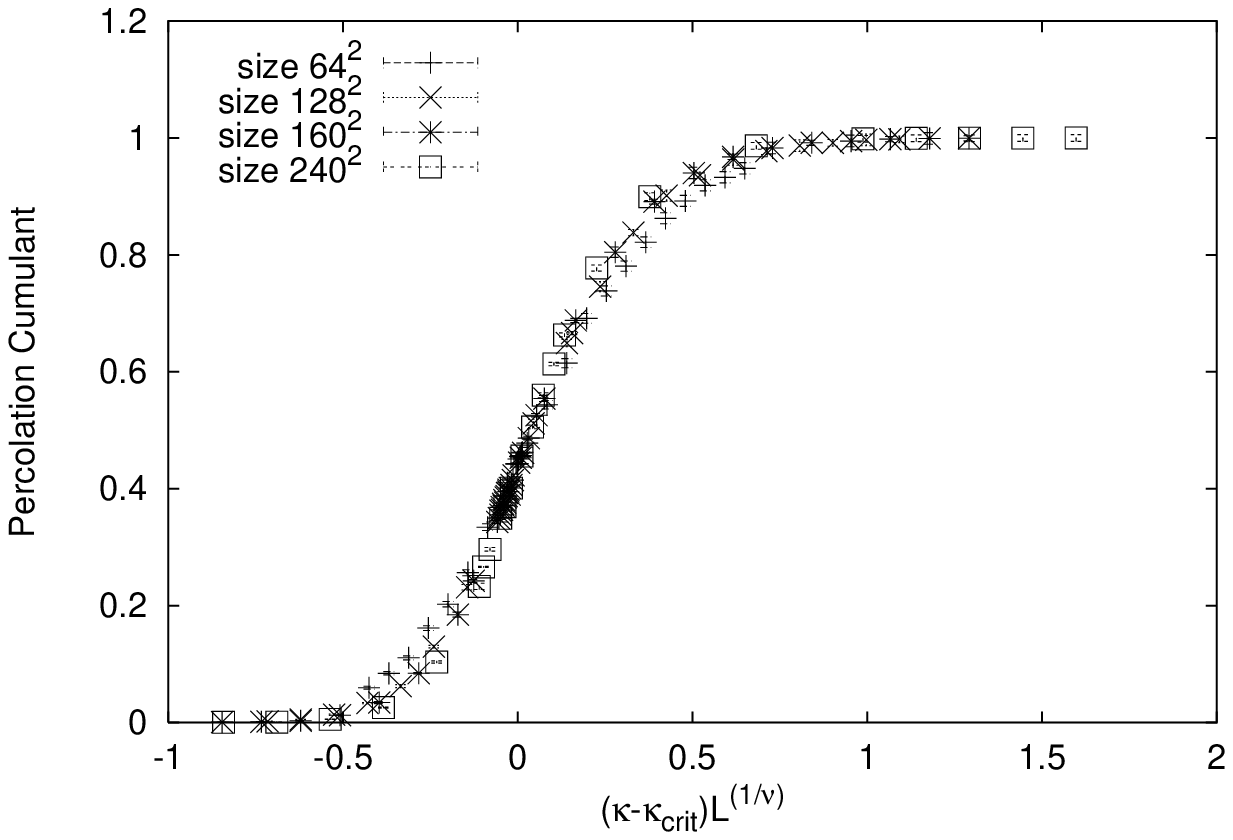,width=9cm}}
    \put(30,-80){\begin{minipage}[t]{12cm}{{\footnotesize
          Figure 4. Rescaled percolation cumulants for Model A
using the 2-dimensional random percolation exponent $\nu=4/3$.}}
\end{minipage}}
\put(80,-280){\epsfig{file=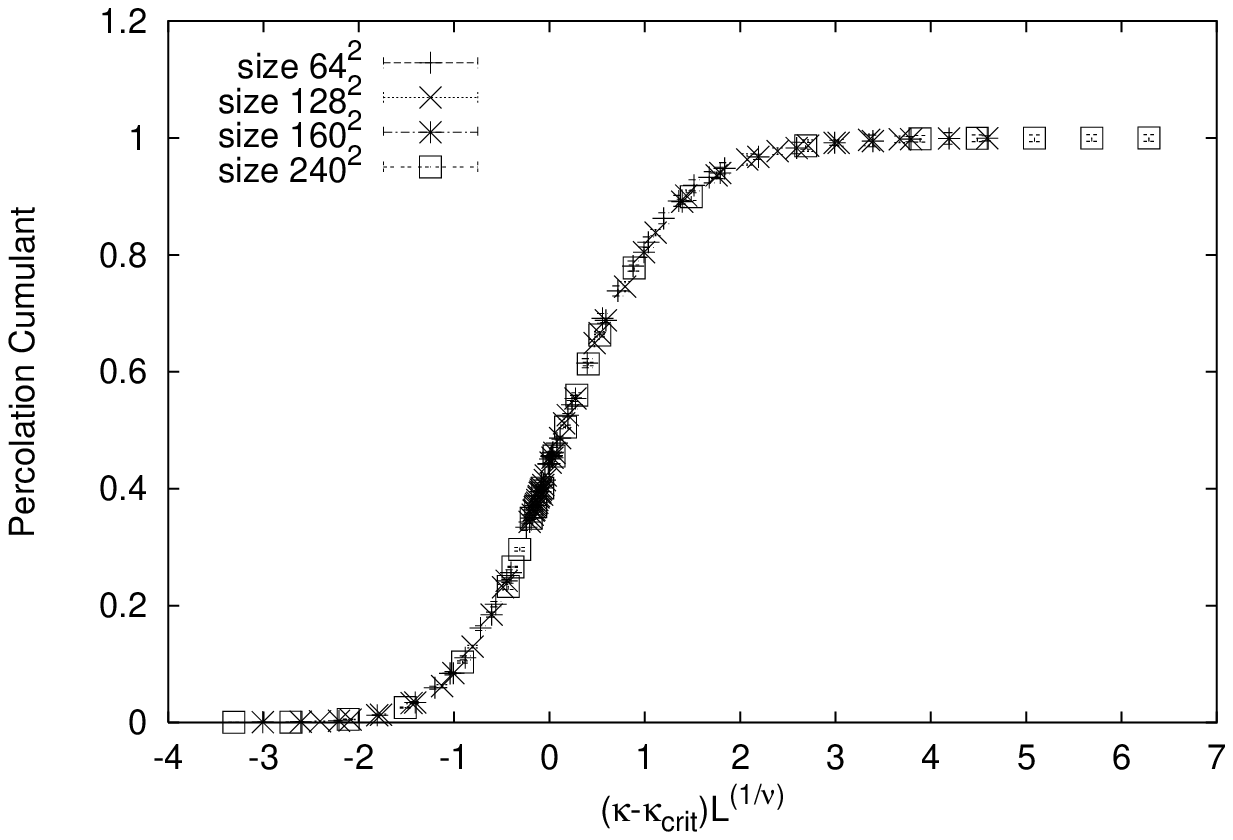,width=9cm}}
    \put(30,-290){\begin{minipage}[t]{12cm}{{\footnotesize
          Figure 5. Rescaled percolation cumulants for Model A
using the 2-dimensional Ising exponent $\nu=1$. }}
\end{minipage}}
\end{picture}

\newpage
\begin{picture}(135,360)
\put(80,140){\epsfig{file=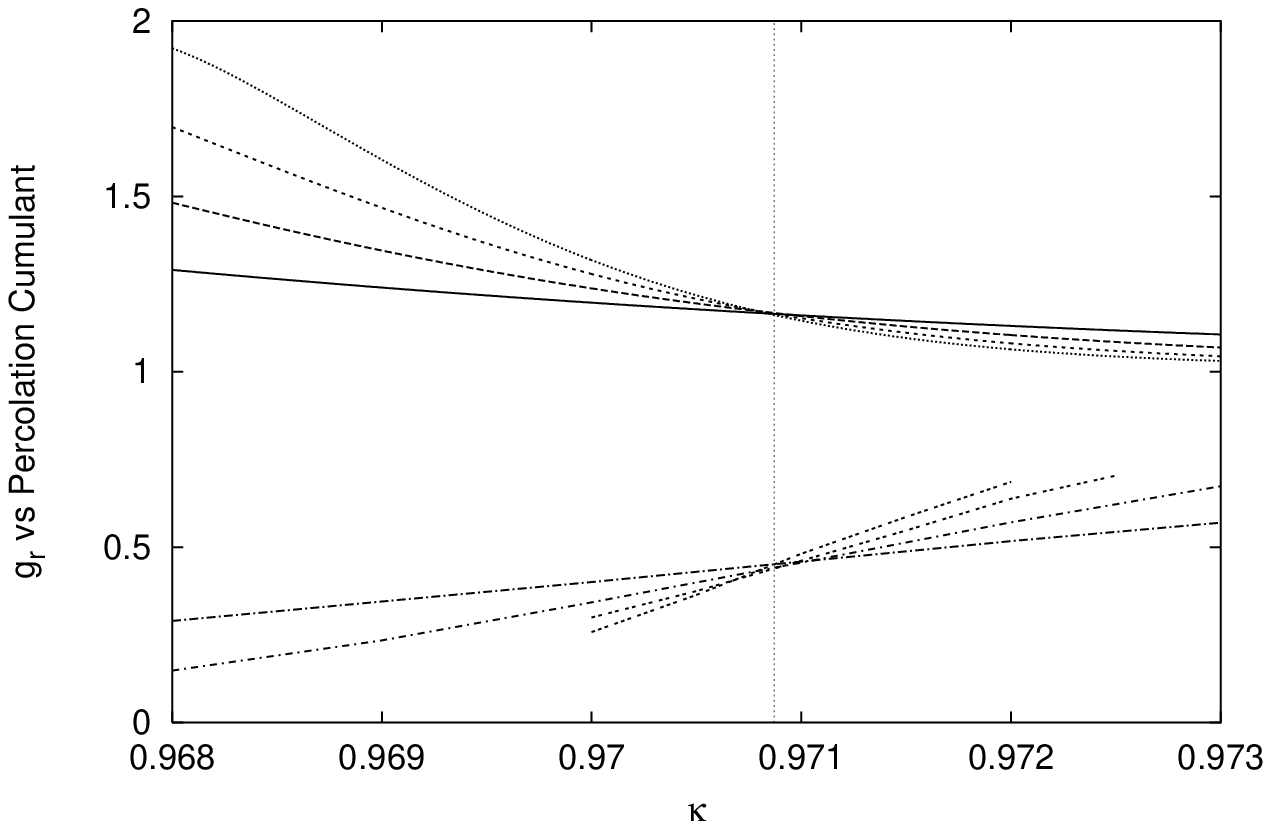,width=9cm}}
    \put(30,130){\begin{minipage}[t]{12cm}{{\footnotesize
          Figure 6. Comparison of the thermal and the geometrical
critical point for Model B obtained respectively from the Binder cumulant
$g_r$ and the percolation cumulant.}}
\end{minipage}}
\put(80,-70){\epsfig{file=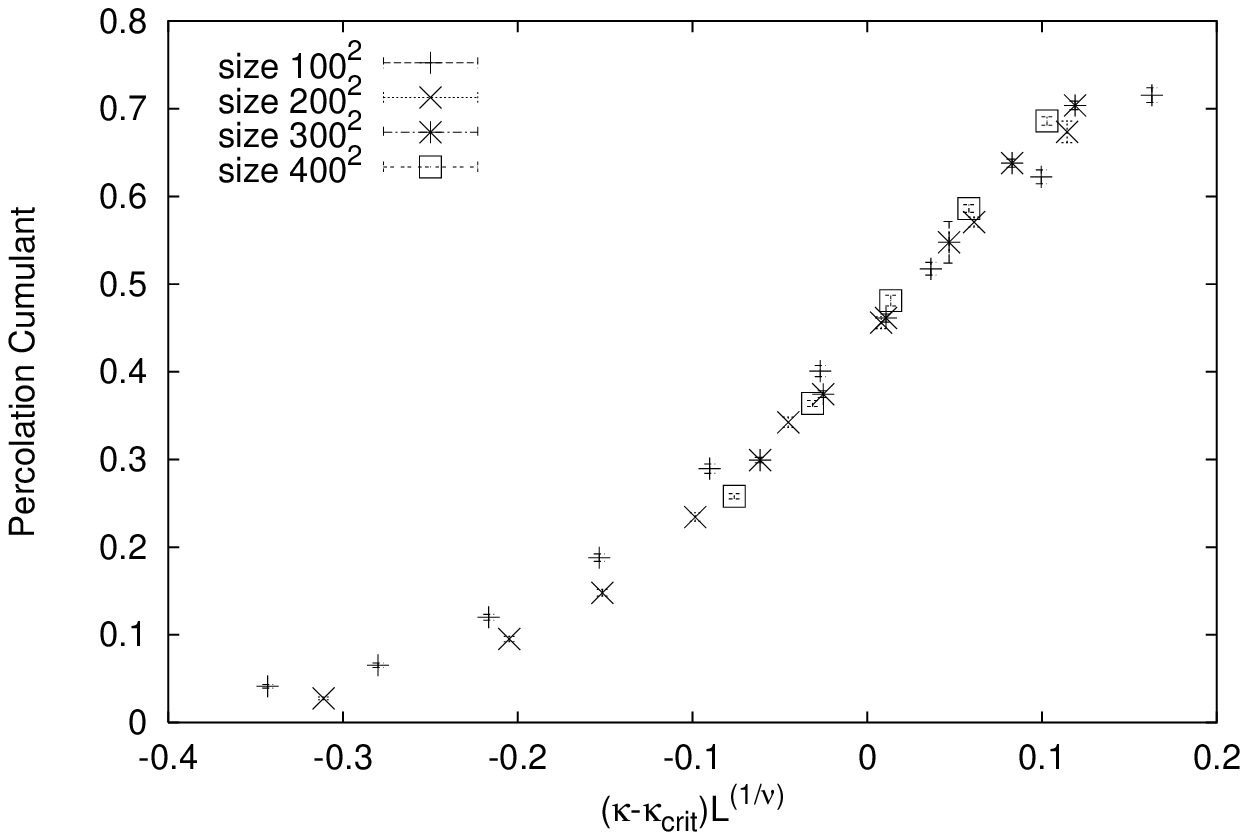,width=9cm}}
    \put(30,-80){\begin{minipage}[t]{12cm}{{\footnotesize
          Figure 7. Rescaled percolation cumulants for Model B
using the 2-dimensional random percolation exponent $\nu=4/3$.}}
\end{minipage}}
\put(80,-280){\epsfig{file=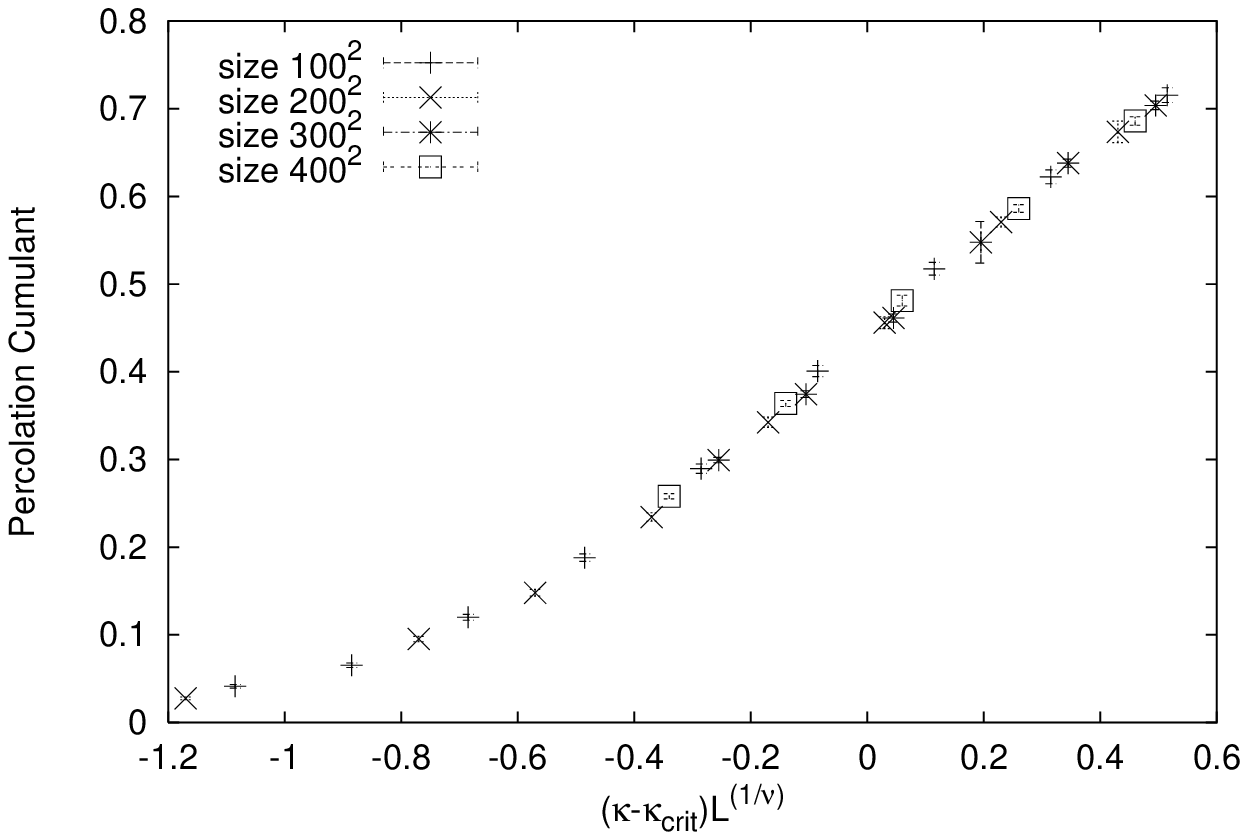,width=9cm}}
    \put(30,-290){\begin{minipage}[t]{12cm}{{\footnotesize
          Figure 8. Rescaled percolation cumulants for Model B
using the 2-dimensional Ising exponent $\nu=1$. }}
\end{minipage}}
\end{picture}

\newpage

\begin{picture}(135,360)
\put(80,140){\epsfig{file=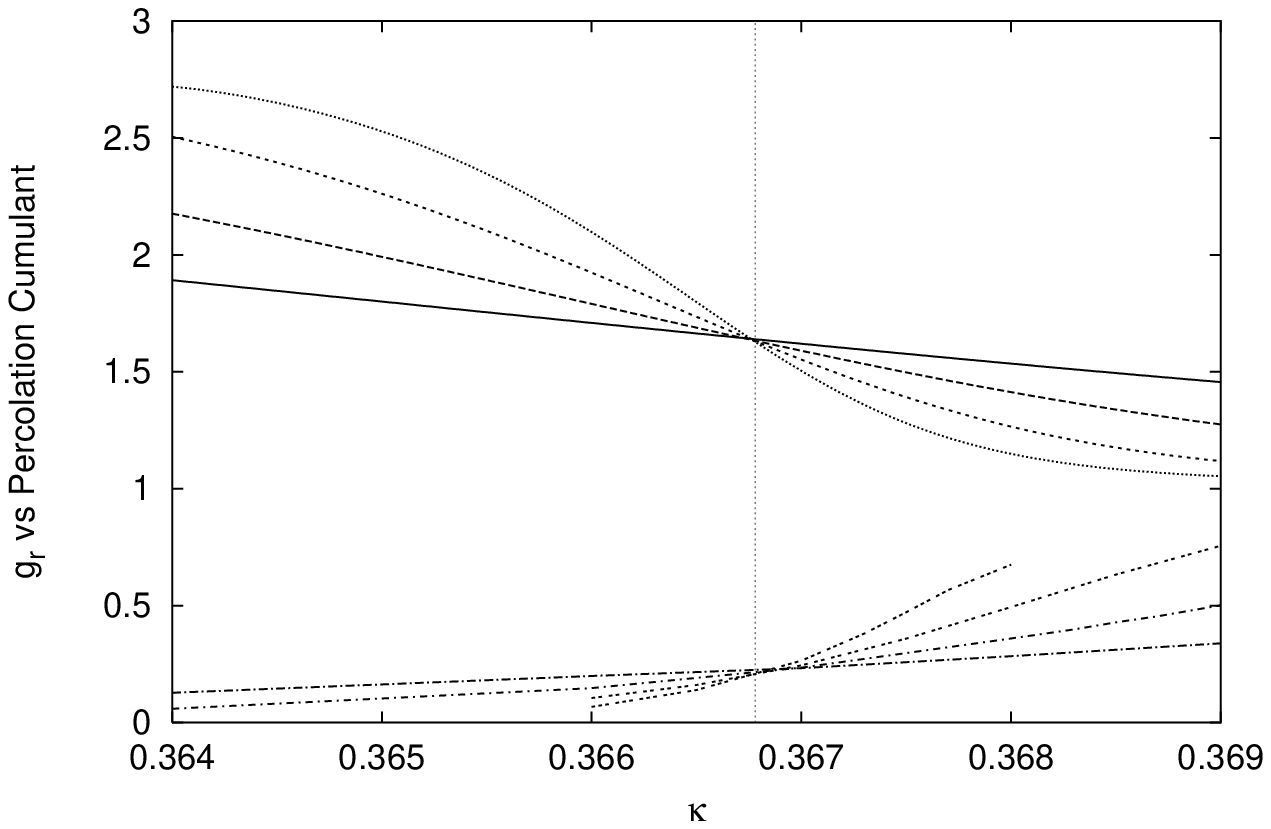,width=9cm}}
    \put(30,130){\begin{minipage}[t]{12cm}{{\footnotesize
          Figure 9. Comparison of the thermal and the geometrical
critical point for Model C obtained respectively from the Binder cumulant
$g_r$ and the percolation cumulant.}}
\end{minipage}}
\put(80,-70){\epsfig{file=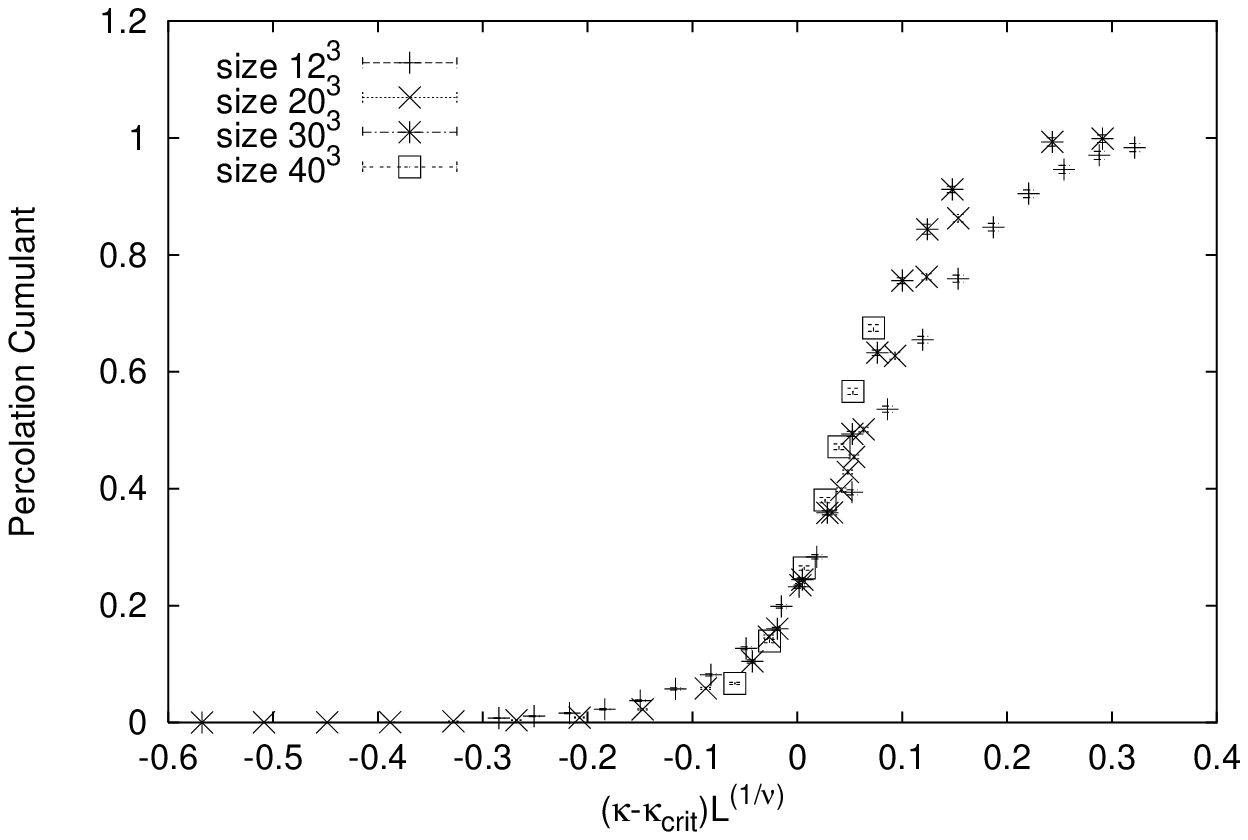,width=9cm}}
    \put(30,-80){\begin{minipage}[t]{12cm}{{\footnotesize
          Figure 10. Rescaled percolation cumulants for Model C
using the 3-dimensional random percolation exponent $\nu=0.88$.}}
\end{minipage}}
\put(80,-280){\epsfig{file=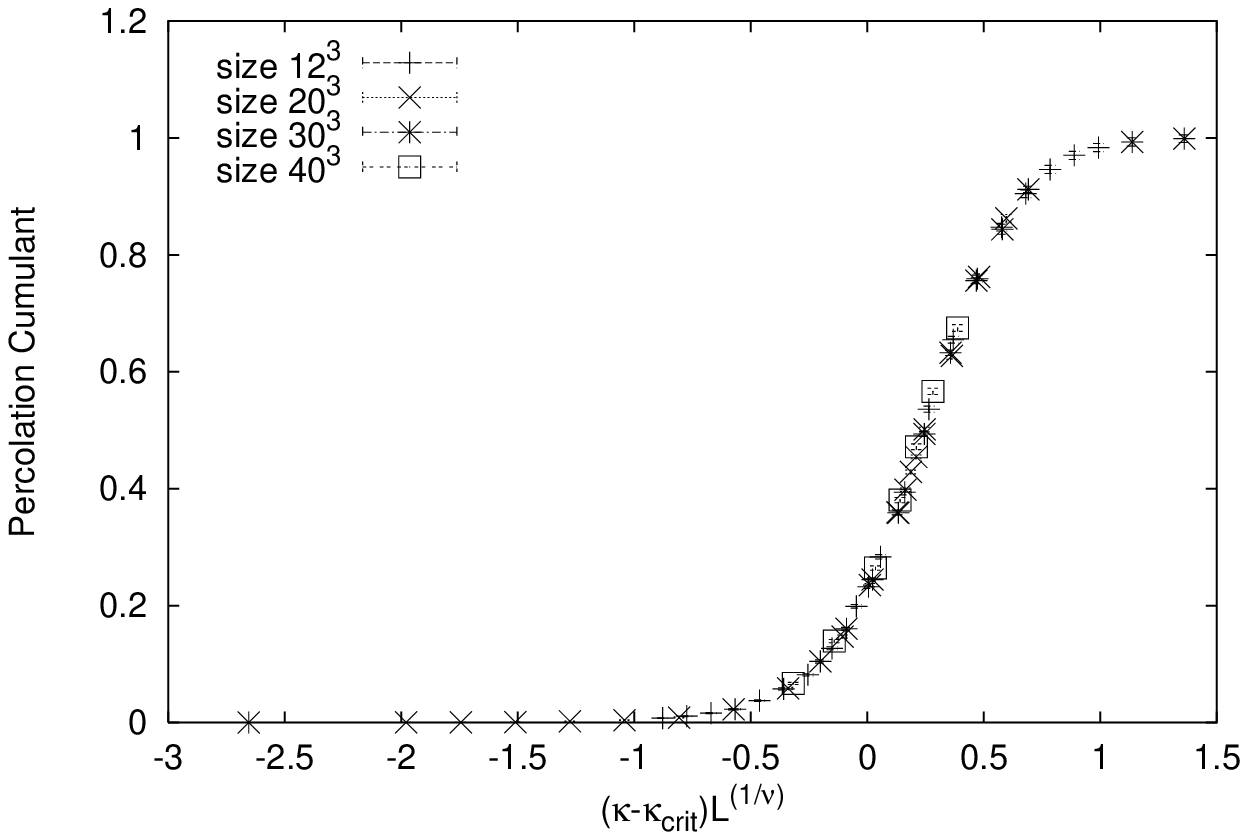,width=9cm}}
    \put(30,-290){\begin{minipage}[t]{12cm}{{\footnotesize
          Figure 11. Rescaled percolation cumulants for Model C
using the 3-dimensional Ising exponent $\nu=0.6289$. }}
\end{minipage}}
\end{picture}

\newpage
\begin{picture}(135,360)
\put(80,140){\epsfig{file=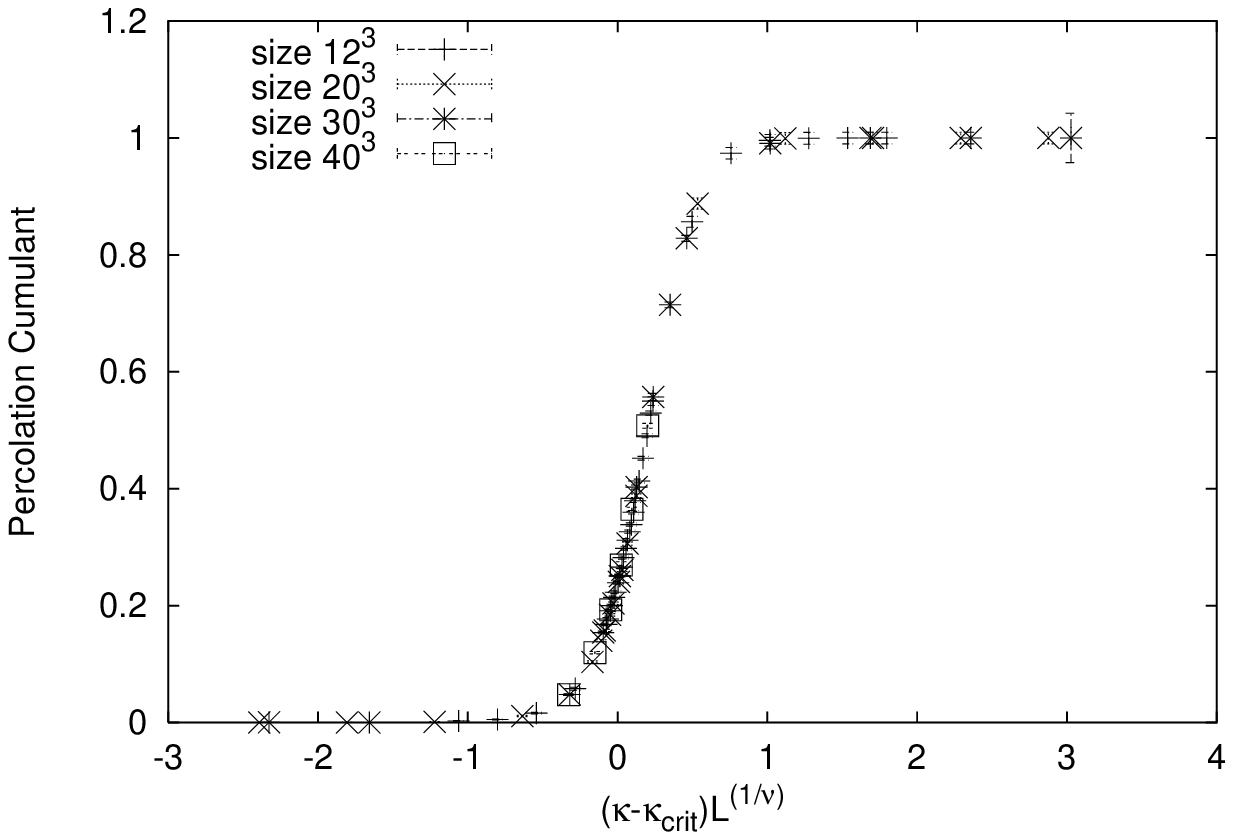,width=9cm}}
    \put(30,130){\begin{minipage}[t]{12cm}{{\footnotesize
          Figure 12. Rescaled percolation cumulants for Model D
using the 3-dimensional Ising exponent $\nu=0.6289$.}}
\end{minipage}}
\end{picture}

\end{document}